\documentclass[10pt,conference]{IEEEtran}
\IEEEoverridecommandlockouts
\usepackage{cite}
\usepackage{amsmath,amssymb,amsfonts}
\usepackage{graphicx}
\usepackage{textcomp}
\usepackage{xcolor}
\usepackage{threeparttable}
\usepackage{booktabs}
\usepackage{tabularx}
\usepackage{url}
\begin{document}
\title{Decoupling Code Complexity from Newcomer Participation: A Causal Study of AI Coding Agent Adoption in OSS}
\author{\IEEEauthorblockN{Weiwei Xu, Xuanning Cui, Hengzhi Ye, Minghui Zhou\thanks{\IEEEauthorrefmark{2}Minghui Zhou is the corresponding author.}\IEEEauthorrefmark{2}}
\IEEEauthorblockA{\textit{School of Computer Science, Peking University, Beijing, China}\\
\textit{Key Laboratory of High Confidence Software Technologies, Ministry of Education, China}\\
\{xuww, hzye\}@stu.pku.edu.cn, \{cxn, zhmh\}@pku.edu.cn}
}
\maketitle

\begin{abstract}
Open-source projects depend on a steady inflow of newcomers. A growing concern is that AI coding agents (tools such as Cursor and Claude Code that write code from natural-language instructions) will crowd them out, by absorbing the simple tasks that beginners start with and by making code harder to read. We give this concern a causal answer. 
Using GitHub code search we identify 1{,}888 projects that adopted an agent, signaled by their first commit of a configuration file. We apply difference-in-differences against matched non-adopting controls, restricting the main analysis to the 603 adopters with a genuine pre-adoption period. We find \emph{no evidence of crowding-out}: across estimators newcomer inflow shows no significant decline after adoption (point estimates run from a small increase to, under the most conservative trend specification, a slight and insignificant dip), onboarding and retention are unchanged, and a sparse, correlational beginner-task measure (good-first-issue labels, which we cannot test for parallel trends) shows no decline. The feared mechanism is real but \emph{decoupled}: adoption raises per-function code complexity (about $+11\%$ on a cognitive metric for Python, a quarter of the prior estimate, and $+3$ to $4\%$ in cyclomatic terms across all languages), yet in fixed-unit subsets where complexity rose (Python on the cognitive metric, and all languages on the cyclomatic metric), newcomer participation does not decline. 
These results suggest that, in established open-source projects, adopting an AI coding agent makes code modestly more complex but does not crowd out the human newcomers that a project depends on: the feared trade-off between AI assistance and human participation does not materialize.
\end{abstract}


\section{Introduction}

Open-source software (OSS) runs on a continual supply of new people. Volunteers come and go, so every project depends on a steady stream of \emph{newcomers} (people making their first contribution) to replace those who drift away and to grow over time~\cite{steinmacher2015social,zhou2015who,xiao2023early}. The early barriers newcomers face, from finding an approachable task to getting a first patch reviewed, are well documented~\cite{steinmacher2015social,steinmacher2015systematic} and shape whether a project stays healthy and survives long-term~\cite{xiao2023early,hao2026first,calefato2022will,coelho2017why}.

Into this setting arrived the AI coding agent. Tools such as Cursor and Claude Code write and edit code directly from plain-language instructions, and a growing body of evidence shows they substantially boost developer productivity~\cite{peng2023impact,ziegler2022productivity,liang2024survey}.

This raises a two-part concern for the newcomer pipeline. First, agents are well suited to exactly the small, self-contained ``good first issues'' newcomers use to make a first contribution~\cite{tan2020first,xiao2022recommending}; if agents absorb that work, the supply of beginner-friendly tasks may narrow. Second, agents may make the code itself harder to approach, since AI assistants affect not just how fast code is written but its quality and structure~\cite{he2025speed,dakhel2023github,pearce2022asleep}. A recent large study finds adopting Cursor raised the \emph{cognitive complexity} of code, that is, how hard it is to read, growing with nesting and tangled control flow, by roughly 41\%~\cite{he2025speed}, and harder-to-read code is harder for a newcomer to understand and modify~\cite{steinmacher2015social}. Together these suggest a plausible mechanism by which AI agents could \emph{crowd out} human newcomers: the easy tasks vanish, and the remaining code becomes harder to read.

Whether this mechanism actually plays out is harder to answer than it first appears. The difficulty is one of causal attribution: even if projects that adopt an agent go on to attract fewer newcomers, that correlation alone would not show the agent caused the decline, 
because the two groups differ before adoption ever occurs~\cite{kalliamvakou2014promises}. Adopters tend to be more active, faster-growing, and more visible, and those same traits independently affect how many newcomers arrive~\cite{zhou2015who,xiao2023early}. A naive comparison would confuse the effect of the agent with the effect of being the kind of project that adopts one, so isolating the agent's own contribution requires a design that nets out these differences~\cite{callaway2021did,borusyak2024revisiting,athey2017state}.

We take such a design to the data. Using GitHub code search we identify 1{,}888 projects that adopted an agent by committing a \texttt{.cursorrules} or \texttt{CLAUDE.md} file, a \emph{visible} adoption signal that is public and datable rather than a measure of usage intensity, mindful of the pitfalls of mining GitHub at scale~\cite{kalliamvakou2014promises,gousios2013ghtorrent,dabbish2012social}. For each we build a comparison group of similar non-adopting projects, matched on pre-adoption age, popularity, language, and recent activity, and apply \emph{difference-in-differences} (DiD)~\cite{callaway2021did}: the before-versus-after change in newcomer inflow in adopters minus the same change in matched controls is the agent's effect. DiD requires a real before-period in which the two groups move together, so our main analysis restricts to an \emph{identified subsample} of 603 adopting projects with at least six months of pre-adoption history, compared against 1{,}784 never-adopter controls selected by the matching stage. On these we estimate the effect on newcomer inflow, how first contributions are handled, whether newcomers stay, how many beginner tasks are opened, and, crucially, code complexity. We check that before-period trends run in parallel before trusting an estimate. Because agents are adopted at \emph{different times} (a staggered-adoption setting in which the textbook two-way fixed-effects estimator is biased~\cite{dechaisemartin2020twoway}), we use several modern estimators~\cite{borusyak2024revisiting,callaway2021did,sun2021estimating} so no result hangs on a single method (Section~\ref{sec:estimation}).

Our central finding is a \emph{measured decoupling}: the feared complexity barrier is real but, at the magnitudes we observe, does not translate into a detectable decline in newcomer participation. Agent adoption modestly raises code complexity (cyclomatic by only about $3$ to $4\%$ across all languages and cognitive by about $11\%$ on Python, roughly a quarter of the prior estimate~\cite{he2025speed}), yet on the repositories where complexity rose newcomer inflow does not decline (point estimates range from flat to mildly positive), and a sparse, correlational measure shows no fall in beginner-task supply. Our work provides a practical insight: for an established project, adopting an agent makes its code modestly more complex but does not crowd out the human newcomers it depends on. We make three contributions.

\begin{itemize}
  \item \textbf{The first causal estimate, to our knowledge, of agent adoption on the OSS newcomer pipeline.} Across newcomer inflow, first-contribution handling, retention, and beginner-task supply, we find no evidence that visible agent adoption crowds out human newcomers in established projects.
  \item \textbf{A field calibration of the complexity cost of agent adoption.} We reproduce the direction of prior evidence that AI-assisted development raises code complexity~\cite{he2025speed}, but estimate much smaller effects in real repositories: roughly $3$--$4\%$ for cyclomatic complexity across languages and about $11\%$ for cognitive complexity on Python.
  \item \textbf{A fixed-unit test of whether the complexity mechanism translates into human exclusion.} On the same projects whose code becomes more complex, newcomer inflow, first-PR activity, and retention do not decline. This shows that the measured complexity increase is real but \emph{decoupled} from the newcomer outcomes it was expected to harm.
\end{itemize}

The remainder of the paper details our data and identification strategy, presents the newcomer and mechanism results, and discusses what they imply for OSS projects and for future studies of AI adoption.

\section{Related Work}

Our study joins three largely separate research lines: what AI coding assistants do to code and developers, how newcomers join and stay in open source, and the methods for credible causal claims in mining software repositories.

\subsection{Effects of AI coding assistants}
A fast-growing literature measures what happens when developers adopt AI coding assistants. The \emph{productivity} strand reports a consistent result, namely faster tasks and more code shipped, in controlled experiments~\cite{peng2023impact,ziegler2022productivity} and grounded studies of real use~\cite{liang2024survey,barke2023grounded}. The \emph{quality} strand is more contested: faster output can bring correctness slips, security weaknesses, and tangled control flow~\cite{dakhel2023github,nguyen2022empirical,pearce2022asleep}. Disentangling the two requires measuring the code itself, a point the broader LLM-for-software-engineering surveys also stress~\cite{fan2023large,hou2024large}.

The closest study to ours, He et al.~\cite{he2025speed}, frames our central question: across a large repository sample, they find adopting Cursor raised the \emph{cognitive complexity} of code (how hard it is to read, rising with nesting and tangled control flow) by about 41\%. Their design is closely related to ours: they date adoption by the first \texttt{.cursorrules} commit and compare adopters against matched non-adopters with the same staggered-adoption imputation estimator, so our magnitudes are closely comparable, although differences in corpus, filtering, pre-window, and metric implementation may still contribute. If AI makes code harder to understand, it may raise a ``complexity barrier'' falling hardest on first-time contributors. He et al. establish the \emph{code-level} channel; whether it translates into \emph{human} exclusion has not been tested, and that is the gap we fill.
\vspace{-1mm}
\subsection{Newcomers and onboarding in open source}
Steinmacher et al.~\cite{steinmacher2015social,steinmacher2015systematic} anchor this literature, cataloguing the barriers a \emph{newcomer} (a first-time contributor) faces: social (unwelcoming reactions), process (unclear guidelines), and, centrally for us, \emph{technical} barriers, including code that is too difficult to understand. Efforts to lower them, from onboarding portals to mentoring, target these obstacles~\cite{steinmacher2016overcoming,fagerholm2014role,balali2018newcomers}.

A parallel strand studies how outsiders become contributors: the joining and socialization paths they follow~\cite{vonkrogh2003community,ducheneaut2005socialization}, the first interactions that shape whether they continue~\cite{jensen2011joining}, and the social links that ease entry~\cite{casalnuovo2015developer}; the common thread is that joining is easily derailed by friction~\cite{steinmacher2015systematic}. Technical barriers are exactly what a complexity-raising tool might worsen, the mechanism behind our hypothesis.
\vspace{-1mm}
\subsection{Newcomer retention and project sustainability}
Converting first-time contributors into lasting ones is what sustains a project, and \emph{early} participation strongly predicts long-term outcomes: a developer's first-contribution and early-window behavior forecasts whether they stay active months or years later~\cite{xiao2023early,zhou2015who}, and the conditions met at entry shape the whole trajectory~\cite{zhou2011does}.

The newcomer's \emph{first task} is a decisive lever: its difficulty, scope, and framing drive whether a contributor stays~\cite{hao2026first}, which is why projects curate ``good first issues''~\cite{tan2020first} and recommend approachable first tasks~\cite{xiao2022recommending}. Beyond it, social capital sustains participation~\cite{qiu2019going} while disengagement is common and only partly reversible~\cite{calefato2022will,lin2017developer}; at the project level, ``magnetism'' and ``stickiness''~\cite{yamashita2014magnet} and ecosystem conditions determine sustained inflow~\cite{valiev2018ecosystem,midha2012factors,zhou2016inflow}, and retention failures propagate into abandonment and project death~\cite{coelho2017why,avelino2019abandonment,foucault2015impact,mockus2002two,constantinou2017empirical}.

Because retention is fragile and sensitive to first experiences, a tool that raises code complexity is a plausible threat: harder-to-read code could degrade the pivotal early experience, thinning retention even if raw inflow holds. The literature thus yields three falsifiable predictions under the barrier hypothesis (fewer newcomers, weaker retention, fewer beginner-friendly tasks) that we test together.

\subsection{Positioning}
Our contribution sits at the intersection of these threads: to our knowledge we are the first to link the \emph{complexity channel} He et al.~\cite{he2025speed} document to the \emph{newcomer participation and retention} that Steinmacher et al.~\cite{steinmacher2015social} and the retention literature~\cite{xiao2023early,hao2026first} show is fragile, and to do so causally. We confirm a modest complexity rise, then show that on the repositories where it rises it \emph{decouples} from the human pipeline.

\section{Research Questions}

Our central claim is that adopting an AI coding agent (a tool that generates and edits code from natural-language prompts, building on the code-LLMs that power them~\cite{chen2021evaluating,vaithilingam2022expectation,imai2022copilot}) does not crowd out human newcomers in established open-source projects, even though it makes the code modestly more complex. We unpack this into three questions that follow the causal chain from human outcome to mechanism to linkage.

\begin{itemize}
  \item \textbf{RQ1 (Human outcome).} Does visible agent adoption narrow the OSS newcomer pipeline? We examine newcomer inflow, first-contribution handling, retention, and beginner-friendly task supply.
  \item \textbf{RQ2 (Mechanism).} Does visible agent adoption increase code complexity in the field, and by how much? We estimate cyclomatic complexity across languages and cognitive complexity on Python.
  \item \textbf{RQ3 (Linkage).} When complexity rises, does newcomer participation decline on the same projects? We re-estimate newcomer outcomes on the exact repositories where the complexity mechanism is observed.
\end{itemize}

\noindent These three questions map one-to-one onto our contributions: RQ1 to the human outcome, RQ2 to the field magnitude of the mechanism, and RQ3 to whether the mechanism reaches people. Each is answered with the same difference-in-differences machinery, so the results rest on a common footing.

\section{Data and Study Design}

We describe, in turn, how we detect adoption (Section~\ref{sec:finding}), the corpus and the sample restriction its structure forces (Sections~\ref{sec:descriptives}, \ref{sec:identification-method}), the matched controls and outcomes (Sections~\ref{sec:controls}, \ref{sec:outcomes}), the estimators (Section~\ref{sec:estimation}), the complexity reconstruction (Section~\ref{sec:complexity}), and the released artifacts (Section~\ref{sec:repro}).

\subsection{Finding projects that adopted an agent}
\label{sec:finding}

We identify adoption through a public, datable signal: the first commit adding an agent's \emph{configuration file} (\texttt{.cursorrules} for Cursor, \texttt{CLAUDE.md} for Claude Code), whose date is the project's \emph{adoption date}. This is a measure of \emph{visible adoption}, not of how heavily the agent is used; we return to this distinction when interpreting effects.

Finding these projects at scale is hard because GitHub's code-search API returns at most 1{,}000 results per query. We beat this cap with \emph{adaptive size-partitioning}: we repeatedly split each query by the byte-size of the matching file (e.g.\ 1--100 bytes, 101--200 bytes), narrowing any band that still saturates the limit until all return fewer than the cap; their union recovers the full file set. For each matching repository we walk its git history to the \emph{first} commit introducing the file, fixing the adoption date precisely.

We exclude commits made by bots and by the agents themselves, since they are not human contributions and would contaminate every outcome. Rather than match on easily-spoofed display names, we anchor exclusion to the stable numeric GitHub user identifier of known bot and agent accounts. This yields \textbf{1{,}888 treated projects} that adopted an agent, of which \textbf{1{,}887} have the commit history required to enter the analysis panel (one is dropped for lacking it).

\subsection{Corpus and descriptive statistics}
\label{sec:descriptives}

\noindent\textbf{Language mix.} Among the 1{,}888 treated projects, TypeScript is the most common primary language, followed by Python, then JavaScript, Go, Rust, C++, Java, and PHP, with a long tail. This matters for RQ2: cyclomatic complexity is measured uniformly across these languages, but the cognitive-complexity metric of He et al.~\cite{he2025speed} is Python-only, so the cognitive analysis runs on the Python subset and the cyclomatic analysis spans the whole corpus.

\noindent\textbf{Adoption clusters in 2024--2026.} Agent configuration files are recent: adoption dates cluster in 2024--2026, most in the latest part of the window. This recency is why so many adopters are ``born with the tool.'' A project existing only briefly before such files became common cannot accumulate human-only history before adopting.

\noindent\textbf{Popularity spread.} We require at least 10 stars (filtering abandoned scratch repositories) and organize projects into five star buckets ($[10,30)$ up to $1000+$) that also stratify the matching (Section~\ref{sec:controls}).

\noindent\textbf{Sparse, heavy-tailed outcomes.} Newcomer counts are zero in most project-months and occasionally spike, so we model outcomes on the $\log(1{+}x)$ scale with count-aware estimators. Good-first-issue labels are rarer still (only a small minority of projects ever attach one), making that outcome our most fragile.

\noindent\textbf{Pre-adoption history is bimodal: $1{,}888 \rightarrow 603$.} The most consequential corpus feature is the sharply bimodal distribution of \emph{pre-adoption history}. Nearly two-thirds of the 1{,}888 treated projects adopt at or near their first commit, with fewer than six months of prior history; these ``born-with-the-tool'' projects have essentially no before-period. The remaining \textbf{603} have a substantial before-period (median well over a year), giving our central parallel-trends tests real pre-period data. These 603, with the never-adopter controls selected by the matching stage, form the \emph{identified sample} on which the main analysis runs. This bimodality is what makes the naive full-corpus estimate untrustworthy; we motivate the restriction in Section~\ref{sec:identification-method} and demonstrate its necessity in Section~\ref{sec:identification-check}.

\begin{table}[t]
\centering
\caption{Analysis samples. The harvest yields 1{,}888 adopters; 1{,}887 enter the panel; 603 have a genuine pre-adoption period and form the identified sample. The complexity analysis (RQ2) uses git-cloned subsets.}
\label{tab:samples}
\setlength{\tabcolsep}{4pt}
\begin{tabular}{@{}lrrr@{}}
\toprule
Sample & Treated & Control & Proj.-mo. \\
\midrule
Harvested adopters & 1{,}888 & n/a & n/a \\
Analysis panel & 1{,}887 & 1{,}784 & n/a \\
Identified ($\ge$6\,mo pre-history) & 603 & 1{,}784 & 55{,}666 \\
Complexity, cyclomatic & 581 & 577 & 38{,}086 \\
Complexity, cognitive (Python) & 128 & 407 & 14{,}316 \\
\bottomrule
\end{tabular}
\end{table}

\noindent\textbf{The identified panel.} Table~\ref{tab:samples} summarizes the analysis samples. The main panel keeps the 603 identified treated projects and the 1{,}784 never-adopter controls from the matching stage, spanning 55{,}666 project-months. The complexity analysis (RQ2) runs on git-cloned subsets, and the decoupling test (Section~\ref{sec:decoupling-result}) re-uses each complexity-panel unit set (the Python subset and the all-language set) so complexity and newcomer access are compared on identical units.

\begin{figure}[t]
\centering
\includegraphics[width=\columnwidth]{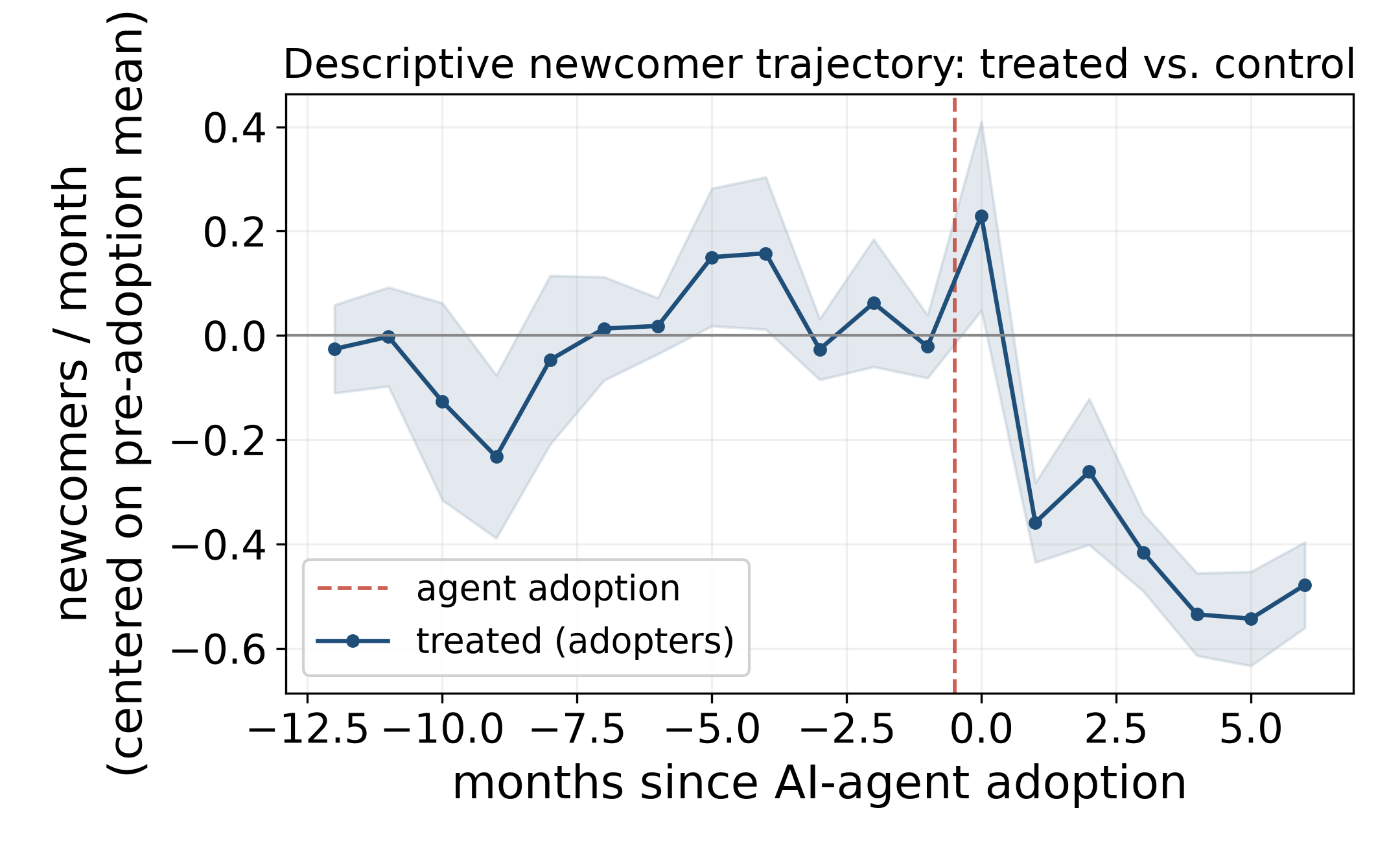}
\vspace{-5mm}
\caption{Descriptive newcomer trajectory in event time, each project centered on its own pre-adoption mean. Treated and calendar-aligned control projects move together before adoption (time 0) and drift down together afterward, a shared calendar trend rather than a treatment effect, which the formal DiD nets out.}
\vspace{-4mm}
\label{fig:descriptive}
\end{figure}

Figure~\ref{fig:descriptive} previews the identification tension: centered on each project's pre-adoption mean, treated inflow tracks calendar-aligned controls \emph{before} adoption, then \emph{both} drift down together after time zero. A naive before/after reads the treated drop as an agent effect; the control trajectory shows it is the shared calendar drift the DiD nets out.

\subsection{Building a comparison group}
\label{sec:controls}

A before-and-after comparison within treated projects alone would confound adoption with everything else changing over time, so we build a \emph{comparison group} of similar non-adopting projects.

\noindent\textbf{The control pool.} Controls are drawn from a \emph{never-adopter} pool (projects that never committed a \texttt{.cursorrules} or \texttt{CLAUDE.md} file as of the crawl); using never-adopters rather than not-yet-adopters keeps the comparison group cleanly untreated, so no control silently becomes treated and contaminates the counterfactual.

\noindent\textbf{Propensity-score matching.} We use propensity-score matching~\cite{austin2011introduction,stuart2010matching}, the \emph{propensity score} being a project's estimated adoption probability given pre-adoption characteristics. We estimate it with a logistic regression (scikit-learn) on \emph{commit-volume} features over the six-month window before adoption (commit count and monthly-commit trend, or slope) alongside age and popularity. Every feature is measured \emph{strictly before} adoption, so the agent cannot have influenced any of them; controls, which have no adoption date, are measured before a language-median pseudo-adoption date.

\noindent\textbf{Star-bucket stratification and 1:3 nearest-neighbor matching.} We stratify by the five star buckets of Section~\ref{sec:descriptives}, matching each treated project only to controls of comparable visibility. Within each stratum we perform \textbf{1:3 nearest-neighbor} matching by propensity score, imposing a \emph{caliper} of $0.2$ times the standard deviation of the logit propensity score~\cite{austin2011introduction} and dropping a treated project whose nearest control lies beyond it. This, plus stratification, is why the control count settles at 1{,}784 rather than exactly three per treated unit.

\noindent\textbf{Balance and an honest caveat.} Table~\ref{tab:balance} reports balance via the standardized mean difference (SMD), the difference in group means scaled by the pooled standard deviation; below $0.1$ is negligible and below $0.25$ acceptable. Stars and forks are well balanced ($|\text{SMD}|\le 0.13$), but three covariates retain mild imbalance: treated projects are somewhat \emph{younger} (age SMD $-0.20$) and more \emph{active}, with higher recent six-month commits ($+0.24$) and a steeper recent commit trend ($+0.29$, the largest difference and slightly above the $0.25$ guideline). Both imbalances run in the same, convenient direction: younger and more active projects tend to attract \emph{more} newcomers, so they would, if anything, inflate treated newcomer counts and work \emph{against} finding crowding-out, making the no-crowding-out conclusion conservative. We do not rest on balance alone: the parallel-trends test reported with each result is the binding check, and the unit-specific-trend specifications (Section~\ref{sec:estimation}) absorb whatever residual gap remains.

\begin{table}[t]
\centering
\caption{Covariate balance after matching, on the never-adopter controls used for the identified sample (603 treated, 1{,}784 control). SMD is the standardized mean difference between treated and control; stars and forks clear $|\text{SMD}|\le 0.13$, while project age and the two activity covariates retain mild imbalance (treated projects are younger and more active), a direction that makes the no-crowding-out finding conservative.}
\label{tab:balance}
\begin{tabular}{lrrr}
\toprule
Covariate & Treated & Control & SMD \\
\midrule
Project age (days)      & 799  & 1019 & $-0.20$ \\
Log stars               & 4.48 & 4.73 & $-0.13$ \\
Log forks               & 2.85 & 3.08 & $-0.12$ \\
Recent 6-mo.\ commits   & 333  & 167  & $+0.24$ \\
Recent commit trend     & 8.14 & 1.61 & $+0.29$ \\
\bottomrule
\end{tabular}
\vspace{-3mm}
\end{table}

\subsection{Outcome measures}
\label{sec:outcomes}

We organize all data into a \emph{monthly panel}: for every project-month a row of outcomes, fine enough to place the adoption date sharply yet coarse enough to smooth day-to-day noise.

A \textbf{newcomer} is a person making their first commit to a project's default-branch history within our window (we treat first-seen-since-2022 as ``first''). The newcomer-inflow outcome is the \emph{newcomer count} per project-month. For onboarding we follow each newcomer's first pull request and record three outcomes from the pull-based-development literature~\cite{gousios2014exploratory,tsay2014influence}: whether it was accepted (merged), time to merge, and time to first response (the first comment or review on the pull request). For retention we record, per newcomer, whether they made any further commit within 3 and within 6 months. For beginner-task availability we count \emph{good-first-issues} (the canonical \texttt{good first issue} label) opened and resolved per month. Throughout, bot and agent activity is removed via the numeric-id exclusion of Section~\ref{sec:finding}, so every count reflects human contributors only.

Because count outcomes are often zero and occasionally large, we model them on the $\log(1{+}x)$ scale, which stabilizes the heavy right tail and lets each estimate read as approximately a percentage change. The two retention outcomes and first-PR acceptance are rates (between 0 and 1), kept on the raw scale and reported in \emph{probability points}.

\subsection{Estimating the effect}
\label{sec:estimation}

\noindent\textbf{Estimand.} We estimate the \emph{average treatment effect on the treated} (ATT) within our \emph{established}-adopter population (at least six months of pre-adoption history). For count outcomes it is the average effect, on the $\log(1{+}x)$ scale, across post-adoption project-months at horizons $h\ge 0$, reported as $100\,(e^{b}-1)$ for a coefficient $b$. The overall ATT averages over all post-adoption months; the event study reports the adoption (configuration-commit) month separately, as it can reflect a mechanical transition rather than a settled state. For the \emph{rate} outcomes (retention at 3 and 6 months and first-PR acceptance) effects are \emph{probability points} (pp): a $+0.057$ retention effect means $5.7$ more newcomers in a hundred remain active.

\noindent\textbf{Primary estimator (BJS imputation).} Our primary estimator is the imputation method of Borusyak, Jaravel, and Spiess~\cite{borusyak2024revisiting}. Using only the comparison projects and the treated projects' \emph{pre-adoption} months, it learns how each outcome would evolve absent adoption, predicts each treated project's counterfactual post-adoption path, and takes the effect as the average actual-minus-predicted gap. Fitting where the agent has no influence and projecting forward, no already-treated observations contaminate the comparison, which is a pitfall of two-way fixed-effects DiD that can flip a true effect's sign under staggered adoption~\cite{goodman2021difference,dechaisemartin2020twoway}.

\noindent\textbf{Confirmatory estimators.} We confirm every headline with two estimators that handle staggered adoption differently. The Callaway--Sant'Anna estimator~\cite{callaway2021did} computes a \emph{group-time} effect per adoption cohort and period and averages them; we report both its \emph{group} aggregate and its \emph{dynamic} (event-time) profile. The Sun--Abraham event study~\cite{sun2021estimating} traces the effect month-by-month relative to adoption. Agreement across three estimators that fail differently guards against any single method's artifact.

\noindent\textbf{Parallel-trends test.} Every estimator rests on \emph{parallel trends}, tested via a Wald test that the pre-adoption leads are jointly zero ($\chi^2$ and $p$ reported with each estimate); a result is trustworthy only when it does \emph{not} reject. It holds on the identified-sample newcomer outcome (Sun--Abraham $\chi^2(5)=10.02$, $p=0.075$) and complexity ($p=0.126$), but is \emph{rejected} on the full corpus. This rejection is itself part of the identification argument, since the restriction in Section~\ref{sec:identification-method} is what makes it pass.

\noindent\textbf{Robustness battery.} We stress every estimate with three checks. (i) \emph{Unit-specific linear trends} give each project its own pre-existing slope; this is our most conservative specification, absorbing gradual drifts matching cannot at the cost of some true effect. (ii) \emph{Placebo adoption dates} move each adoption twelve months earlier; a near-zero placebo supports a causal reading, a large one warns of selection on growth. (iii) The \emph{honest-DiD} sensitivity analysis of Rambachan and Roth~\cite{rambachan2023honest,roth2023whats} asks how large a post-adoption parallel-trends violation (a multiple $\bar{M}$ of the largest pre-adoption deviation) must be before the conclusion flips; we report the breakdown $\bar{M}$ to characterize how the identified-sample interval responds to trend violations (it includes zero throughout but widens) and to quantify how fragile the naive full-sample negative is.

\subsection{A necessary identification step}
\label{sec:identification-method}

DiD requires a \emph{before} period. A project committing its agent configuration in its first weeks has essentially no pre-adoption history, so its counterfactual cannot be estimated and parallel trends cannot be tested. As Section~\ref{sec:descriptives} showed, nearly two-thirds of the 1{,}888 treated projects are ``born with the tool,'' and for them the DiD design is unidentified.

We therefore restrict the main analysis to the \textbf{identified sample}: the \textbf{603 treated projects} with at least six months of genuine pre-adoption history and the \textbf{1{,}784 never-adopter controls} from the matching stage, spanning \textbf{55{,}666 project-months}. This is principled, not a convenience cut: the six-month threshold is the minimum pre-window over which we can estimate a baseline trajectory and run the parallel-trends lead test, and the subset on which the counterfactual is well defined (Table~\ref{tab:balance} is computed on it). Including the unidentified projects does not merely add noise: it injects a composition artifact that the parallel-trends test flags as a rejection (Section~\ref{sec:identification-check}). We report this restriction transparently and examine its consequence in a sensitivity check (Section~\ref{sec:identification-check}). Concretely, crowding-out would show up as a \emph{significant negative} effect on newcomer inflow that survives the parallel-trends test on the identified sample; the absence of any such effect (Section~\ref{sec:rq1}) is what we report.

\subsection{Measuring code complexity}
\label{sec:complexity}

RQ2 requires the complexity of each project's code \emph{as it stood in each month}, since the question is how complexity \emph{evolved} around adoption. For each project-month we rebuild the \emph{month-end snapshot} from git: we read the tree at the last commit of the month (\texttt{git ls-tree}) and fetch each file by blob hash (\texttt{git cat-file}). Because an unchanged file keeps the same blob hash across months, we cache complexity by blob hash, making full-history reconstruction over thousands of repositories tractable.

On each snapshot we score every function with two off-the-shelf tools: \textbf{lizard} gives \emph{cyclomatic complexity}~\cite{mccabe1976complexity,ebert2016cyclomatic} (independent paths through a function) across all major languages, and \textbf{complexipy} gives \emph{cognitive complexity}~\cite{campbell2018cognitive} (readability under nesting and tangled control flow) for Python, the same cognitive-complexity metric family used by He et al.~\cite{he2025speed}, so our magnitudes can be compared against theirs. We exclude vendored, generated, and test code and summarize each snapshot by the \emph{per-function average} rather than a total, so the metric tracks how intricate the code is, not how much of it there is. (Size measures, which conflate the two, fail their pre-trend test and are dropped.)

The resulting panel covers \textbf{581 treated and 577 control} projects (\textbf{38{,}086 project-months}) for cyclomatic complexity across all languages, with a Python subset of \textbf{128 treated and 407 control} projects (about \textbf{14{,}316 project-months}) for the cognitive metric. These feed the same estimators and parallel-trends tests as the human-pipeline outcomes, so the decoupling test (Section~\ref{sec:decoupling-result}) can hold each unit set (Python and all-language) fixed across both.

\subsection{Data and reproducibility}
\label{sec:repro}

We pin the analysis so it can be reproduced from the released code and derived panels. The released harvesting code contains the dated code-search crawl, the adaptive size-partitioning queries, the bot/agent numeric-id exclusion list, and first-commit detection. The shipped panels reproduce the main identified-sample estimates; the full pipeline, including the full-sample sensitivity check, rebuilds from scratch by re-harvest. We pin the toolchain: \textbf{lizard 1.23.0}, \textbf{complexipy 5.6}, and the R packages \textbf{didimputation 0.5.1}, \textbf{did 2.5.0}, \textbf{fixest 0.14.1}, and \textbf{HonestDiD 0.2.8}. Standard errors are clustered at the project level throughout; we discuss limits in Section~\ref{sec:threats}.

\section{Results}

We organize results by research question. The central result is a \emph{measured decoupling}: across every outcome bearing on whether human newcomers are pushed out (inflow, onboarding, retention, and beginner-task supply) we find \emph{no evidence of} crowding-out, while confirming adoption modestly raises code complexity.

Table~\ref{tab:summary} collects the headline effects. Three conventions apply throughout. First, unless noted, every effect is estimated on the \emph{identified subsample} (603 treated, 1{,}784 control; 55{,}666 project-months). Second, count outcomes (newcomers, beginner tasks, complexity) enter on the $\log(1{+}x)$ scale, so a coefficient reads as a percentage change $100\,(e^{b}-1)$. Third, the two retention outcomes and first-PR acceptance are \emph{rates} on a $0$--$1$ scale, so their effects are \emph{probability points} (pp), marked to avoid confusion with percentage effects on counts. Our primary estimator is Borusyak--Jaravel--Spiess (BJS) imputation~\cite{borusyak2024revisiting}, confirmed with Callaway--Sant'Anna group-time and dynamic estimators~\cite{callaway2021did} and the Sun--Abraham event study~\cite{sun2021estimating}. For complexity the table reports the \emph{trend-robust} estimate the prose relies on, with the raw BJS estimate in parentheses, since the raw number partly reflects selection on growth.

\begin{table}[t]
\centering
\caption{Headline effects by research question (identified sample unless noted). Count outcomes use $\log(1+x)$ and read as approximate percentage changes; retention and first-PR acceptance are on the raw $0$--$1$ rate scale and are reported in probability points (pp). For complexity we report the trend-robust estimate the prose relies on, with the raw BJS estimate in parentheses. An asterisk ($^{*}$) marks significance at $5\%$; ``n.s.''~=~not significant. RQ1 shows no evidence of crowding-out: the effect is never significantly negative, though the positive point estimate is not robust to unit-specific trends. A dagger ($\dagger$) marks a \emph{correlate}: the good-first-issue effect rests on only about 92 projects whose pre-trends we could not validate, so it is interpreted as correlational, not causal. The RQ3 decoupling rows report trend-robust estimates.}
\label{tab:summary}
\setlength{\tabcolsep}{4pt}
\begin{tabular}{@{}llr@{}}
\toprule
RQ & Outcome & Effect \\
\midrule
RQ1 & Newcomer inflow / month & $+5.3\%^{*}$ (not trend-robust) \\
 & First-PR acceptance & $+2.7$ pp (n.s.) \\
 & Time-to-merge & $-4\%$ (n.s.) \\
 & Time-to-first-response & $-14\%$ (n.s.) \\
 & Retention @ 6 mo & $+2.2$ pp (n.s.) \\
 & Retention @ 3 mo & $+1.4$ pp (n.s.) \\
 & Good-first-issues opened & $+27\%^{*\dagger}$ \\
\midrule
RQ2 & Cyclomatic compl.\ (all lang.) & $+3.3\%^{*}$ (raw $+4.3\%$) \\
 & Cognitive compl.\ (Python) & $+10.8\%^{*}$ (raw $+18.5\%$) \\
\midrule
RQ3 & Decoupling, Python & \shortstack[r]{inflow $+3.4\%$, PRs $+3.1\%$,\\ ret.\ $+5.7$/$+7.7$ pp (n.s.)} \\
 & Decoupling, all lang. & \shortstack[r]{inflow $+3.3\%$, PRs $+0.6\%$,\\ ret.\ $+4.7$/$+5.7$ pp (n.s.)} \\
\bottomrule
\end{tabular}
\end{table}

\subsection{RQ1: Does adoption narrow the newcomer pipeline?}
\label{sec:rq1}

We examine four facets of newcomer participation: inflow, first-contribution handling, retention, and beginner-friendly task supply. Inflow, onboarding, and retention are our primary evidence; the good-first-issue result is a supportive but sparse correlate.

\noindent\textbf{Inflow.}

The crowding-out hypothesis predicts that after adoption monthly newcomer inflow should fall, as the agent absorbs the small tasks first-timers use to begin contributing. We find no such reduction: across every estimator the effect sits between zero and a mild increase and is \emph{never} significantly negative. BJS gives $+5.3\%$ (se $1.2\%$; $95\%$ CI $+2.9\%$ to $+7.8\%$), Sun--Abraham $+4.0\%$ (se $1.9\%$), and the Callaway--Sant'Anna family $+3.1\%$ (simple), $+3.7\%$ (grouped), and $+2.8\%$ (dynamic). The only specification turning negative is our most aggressive, per-project linear trends: under this detrending the estimate falls to $-1.5\%$ (se $0.9\%$, n.s.), still an order of magnitude from the double-digit drop crowding-out requires. This $-1.5\%$ to $+5.3\%$ spread is too wide to assert a precise positive effect but bounded well away from the predicted negative region, and a placebo returns $+0.0\%$ (n.s.) when each adoption date is moved twelve months earlier.

The identifying assumption holds on this exact panel: the Wald test that the pre-adoption leads ($h=-6$ to $-2$) are jointly zero does \emph{not} reject ($\chi^2(5)=10.02$, $p=0.075$; $p=0.19$ with unit-specific trends). Under honest-DiD~\cite{rambachan2023honest}, the robust $95\%$ interval for the average post-period ATT is $[-0.3\%,\,+6.7\%]$ under parallel trends ($\bar{M}=0$) and continues to include zero as the allowed post-period violation grows, so no significant negative effect is forced at any sensitivity level we examine. The interval does widen as that allowance grows (about $[-10.3\%,\,+18.3\%]$ at $\bar{M}=0.5$), so this sensitivity bound alone cannot rule out a sizeable negative; our case against crowding-out rests instead on the positive point estimates, the clean placebo, and the parallel pre-trends.

\begin{figure}[t]
\centering
\includegraphics[width=\columnwidth]{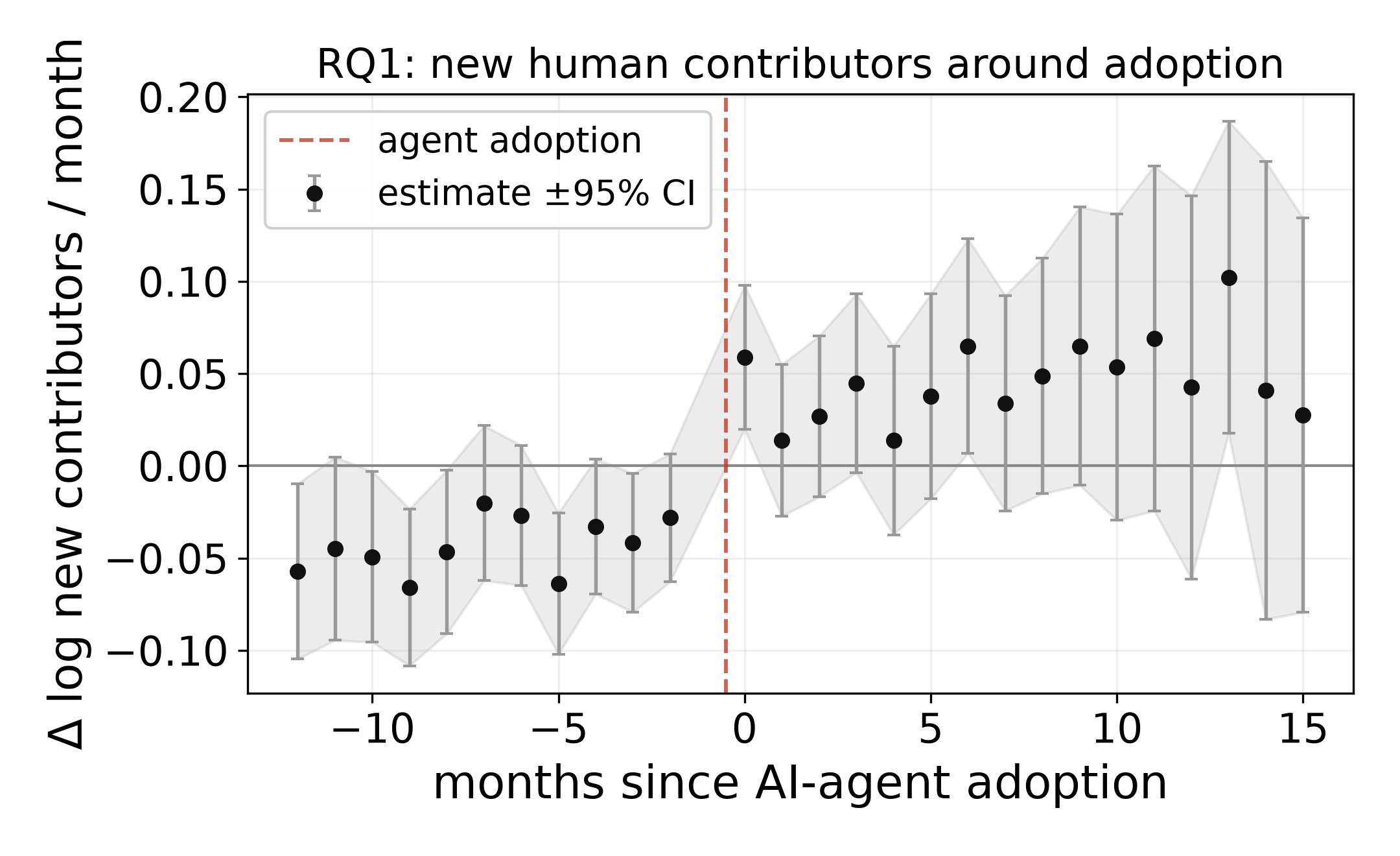}
\caption{RQ1 newcomer event study (identified sample). Monthly newcomer inflow is flat-to-slightly-negative before adoption (time 0), steps up at adoption, and stays elevated for roughly fifteen months. The pre-adoption leads are jointly indistinguishable from zero ($\chi^2(5)=10.02$, $p=0.075$), and there is no decline after adoption.}
\label{fig:newcomers}
\end{figure}

Figure~\ref{fig:newcomers} shows the event study: inflow is flat-to-slightly-negative before adoption, then steps up by roughly $+4$ to $+6\%$ and stays elevated for about fifteen months, with no post-period decline. We therefore find \emph{no evidence} of crowding-out, but stop short of asserting a positive causal effect, since the magnitude is sensitive to trend modeling and adoption may coincide with rising project visibility that independently attracts newcomers. The confident claim is the negative one: there is no evidence adoption pushes human newcomers out.

\noindent\textbf{Onboarding.}

We ask whether adoption changes how a newcomer's first contribution is \emph{handled} (an evaluation outcome shaped by many social and technical factors~\cite{tsay2014influence,terrell2017gender,rastogi2018relationship}) even if how many arrive is unchanged.

We track three outcomes on the subset of project-months with at least one first PR ($n\approx 9{,}500$ across 519 treated projects). All three are null: First-PR \emph{acceptance} (merge) changes by $+2.7$ probability points (i.e.\ $+0.027$ on the $0$--$1$ rate scale, se $1.9$ pp, n.s.), median \emph{time-to-merge} by $-4\%$ (se $0.12$, n.s.), and \emph{time-to-first-response} by $-14\%$ (se $0.11$, n.s.), the last two on the $\log(1{+}x)$ scale. Callaway--Sant'Anna agrees in sign on acceptance (also positive) and is likewise insignificant, though it differs in sign on time-to-merge, underscoring that none of these onboarding effects is robustly distinguishable from zero. Onboarding is thus null, precisely so for acceptance and more loosely for the noisier timing outcomes (Section~\ref{sec:threats}); the inflow result is not masking a deterioration in how first contributions are treated.

\noindent\textbf{Retention.}

We next ask whether newcomers \emph{stay}: a tool could leave inflow and onboarding untouched yet thin a project's future by driving away first-timers before they become regulars. For each newcomer we record any further commit within three and within six months.

\begin{figure}[t]
\centering
\includegraphics[width=\columnwidth]{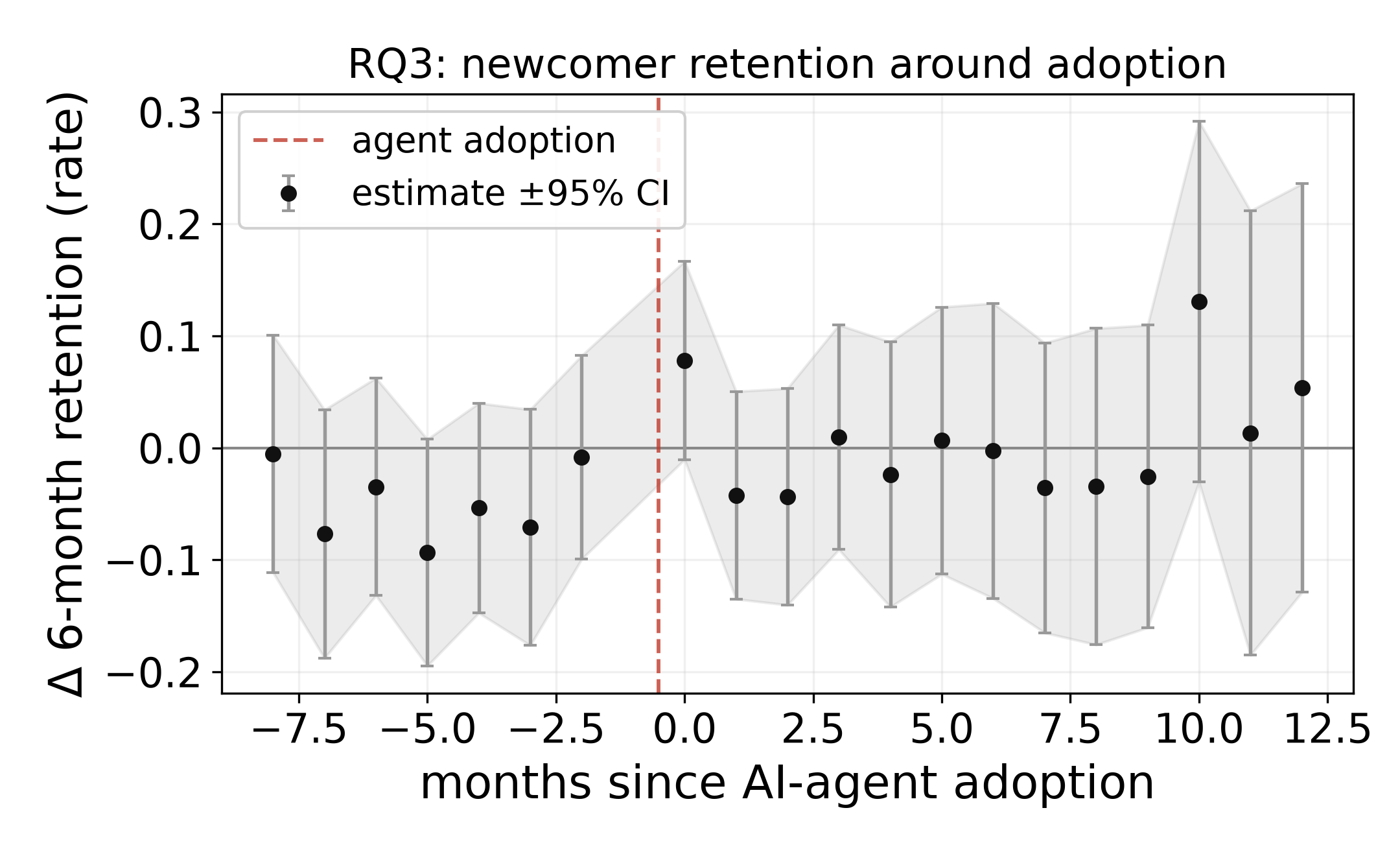}
\caption{Newcomer retention event study (six-month retention, raw $0$--$1$ rate). Pre-adoption leads scatter around zero (parallel trends hold, $p=0.34$), and post-adoption horizons hover near zero with wide intervals, showing no decline in whether newcomers stay.}
\label{fig:retention}
\end{figure}

Both effects are null. Under BJS, retention at six months is $+2.2$ pp (se $1.7$ pp, n.s.) and at three months $+1.4$ pp (se $1.7$ pp, n.s.). Callaway--Sant'Anna sits on the other side of zero but is equally insignificant ($-0.5$ pp at six months, $-1.2$ pp at three, both grouped), and parallel trends hold ($\chi^2(5)=5.7$, $p=0.34$). Retention is our noisiest outcome, with wide standard errors and right-window truncation giving recent newcomers little time to demonstrate staying, so we read it as ``no large effect.'' The central estimate is mildly \emph{positive}, and nothing suggests adoption thins the cohort who remain. Figure~\ref{fig:retention} shows the event study; pre-adoption leads scatter around zero and post-adoption horizons hover near zero with wide intervals, a null rather than a decline.

\noindent\textbf{Beginner-friendly tasks.}
\label{sec:rq4}

Do projects open fewer beginner tasks after adopting? This is the most direct test of the mechanism's first half, that agents absorb the easy work newcomers depend on. We measure entry points by \emph{good-first-issues} (GFIs): tasks a maintainer labels as suitable for a new contributor.

GFIs \emph{opened} per month \emph{rise} by $+27\%$ after adoption (BJS, se $9.6\%$, significant; Callaway--Sant'Anna agrees at $+26\%$), which points against crowding-out: adopting projects advertise \emph{more} beginner work, not less. Downstream GFI outcomes (resolved, bot-closed) are null, so the effect is on the supply of labeled beginner tasks rather than their disposition.

This is the most fragile result in the paper. GFI labels are sparse: only \emph{92} treated projects in the identified sample ($n=948$ project-months) ever carry one, and we could \emph{not} validate its parallel-trends assumption (too few pre-adoption GFI observations survive the joint lead test), so, unlike the inflow and complexity outcomes, we cannot demonstrate parallel before-period trends. We therefore present the GFI rise as a \emph{correlate} consistent with the no-crowding-out result, not an established causal effect.

\subsection{RQ2: Code complexity in the field}

The feared mechanism is code complexity: if agents make code structurally harder to read, a newcomer should find it harder to understand and modify. We test that first link, whether adoption raises complexity, and confirm it while showing the rise is far smaller than the strongest prior estimate, using both \emph{cyclomatic} (independent paths~\cite{mccabe1976complexity}) and \emph{cognitive} complexity (readability under nesting~\cite{campbell2018cognitive}, a comparable metric to He et al.~\cite{he2025speed}).

Cyclomatic complexity across all languages (581 treated, 577 control; 38{,}086 project-months) rises $+4.3\%$ (se $1.0\%$, $p<.001$), with parallel trends \emph{holding} (Wald $p=0.126$). It survives the unit-specific-trend specification at $+3.3\%$, and the twelve-month-early placebo returns $+1.6\%$ ($p=.026$), showing part of the raw rise is selection on growth that the trend-robust estimate nets out. The fraction of individually high-complexity functions rises only $+0.6$ percentage points, so the shift is broad and gradual. Cognitive complexity on the Python subset (128 treated, 407 control; about 14{,}316 project-months) rises $+18.5\%$ raw and $+10.8\%$ ($p=.002$) under unit-specific trends; here the placebo is \emph{cleaner} at $+6.8\%$ but no longer significant ($p=.062$), strengthening the causal reading. Cyclomatic complexity on those same Python repositories rises $+8.1\%$ (trend-robust), confirming the effect is not a metric artifact. Table~\ref{tab:complexity} collects these figures.

\begin{table}[t]
\centering
\caption{Complexity effects: Raw, trend-robust, and placebo estimates.}
\label{tab:complexity}
\setlength{\tabcolsep}{4pt}
\begin{tabular}{@{}lccc@{}}
\toprule
Metric & Raw & Trend-rob. & Placebo \\
\midrule
Cyclomatic (all lang.) & $+4.3\%$ & $+3.3\%$ & $+1.6\%$ ($p{=}.026$) \\
Cognitive (Python)     & $+18.5\%$ & $+10.8\%$ & $+6.8\%$ (n.s.) \\
Cyclomatic (Python)    & n/a       & $+8.1\%$  & n/a  \\
\bottomrule
\end{tabular}
\begin{tablenotes}
\footnotesize 
\item[*] 
Cyclomatic is measured across all languages; cognitive is Python only (\texttt{complexipy}, a comparable cognitive-complexity metric to He et al.~\cite{he2025speed}). ``Trend-robust'' adds unit-specific linear trends; ``placebo'' moves the adoption date twelve months early. Parallel trends hold for cyclomatic complexity ($p=0.126$); the cleaner (n.s.) cognitive placebo strengthens the causal reading.
\end{tablenotes}
\end{table}

\begin{figure}[t]
\centering
\includegraphics[width=\columnwidth]{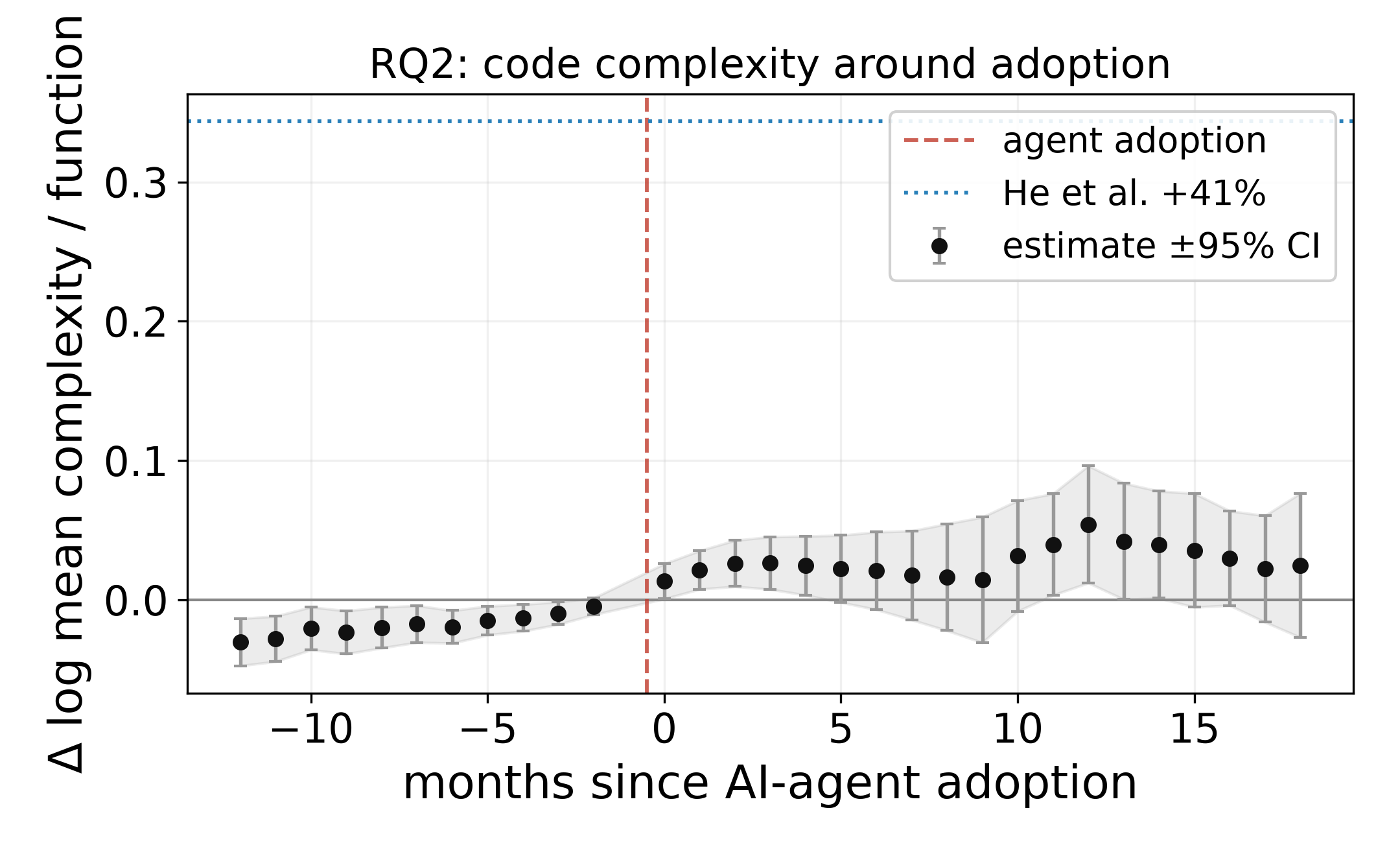}
\caption{RQ2 complexity event study. Code complexity is flat before adoption (time 0), steps up modestly afterward, and settles a few percent higher, far below the $+41\%$ benchmark (dotted line) reported by He et al.~\cite{he2025speed}. Our trend-robust estimates are roughly one-quarter (cognitive) to one-tenth (cyclomatic) of that benchmark.}
\label{fig:complexity}
\end{figure}

Figure~\ref{fig:complexity} shows the event study: flat before adoption, a modest step up afterward, settling a few percent higher, far below the $+41\%$ benchmark (dotted line) He et al.~\cite{he2025speed} report. Against it our cognitive estimate ($+10.8\%$) is about one-quarter and our cyclomatic estimate ($+3.3\%$) about one-tenth; the channel \emph{replicates in direction} but is far weaker in our broad, established-project sample than in the prior, Cursor-only one. We deliberately do \emph{not} report raw \emph{size} measures (total complexity, lines of code, function count), which jump $+74$ to $+86\%$ but whose pre-adoption trends are clearly \emph{violated} ($p\le.001$), marking them as selection (adopting projects are in a fast-growing phase) rather than a causal effect; we interpret only the per-function metrics, whose parallel trends we can defend.

\subsection{RQ3: Does the complexity rise reach newcomers?}
\label{sec:decoupling-result}

One gap remains. The complexity rise (RQ2) and the flat-to-positive newcomer outcomes (RQ1) so far rest on \emph{overlapping but not identical} project sets, leaving open that the two effects occur on different repositories, which would make the decoupling an artifact of aggregation. We address this by holding the unit set \emph{fixed}.

We take the exact 128 treated and 407 control Python projects whose cognitive complexity rose $+10.8\%$ (trend-robust, $p=.002$) and re-estimate every human-access outcome on that identical set. On these projects, whose code became roughly $11\%$ more cognitively complex, newcomer inflow is $+5.7\%$ raw and $+3.4\%$ under unit-specific trends (n.s.); the remaining outcomes, all reported trend-robust, are likewise null or positive: retention $+5.7$ and $+7.7$ probability points at three and six months (both n.s.), newcomer pull-request volume $+3.1\%$ (n.s.), and the active-contributor base $+22.8\%$ (significantly positive). Every human-access outcome is null or positive on the precise projects whose code grew harder to read.

The decoupling is not specific to Python. On the all-language cyclomatic panel (581 treated, 577 control) whose complexity rose $+3.3\%$, the same pattern holds (estimates trend-robust unless noted): inflow $+5.2\%$ raw and $+3.3\%$ trend-robust, retention $+4.7$ and $+5.7$ pp at three and six months (both n.s.), newcomer pull-request volume $+0.6\%$ (n.s.), and the active-contributor base $+20.0\%$, with parallel trends holding ($p=0.086$) and a clean placebo ($+0.5\%$). Across both the cognitive (Python) and cyclomatic (all-language) panels, complexity rose while newcomer access did not fall.

This is the clearest statement of the thesis: unit for unit, complexity rose while newcomer access did not decline. The barrier prior work feared is real and measurable, but on established open-source projects it is modest and decoupled enough from human participation that it does not crowd out newcomers.

\subsection{Sensitivity check: the full sample lacks a valid before-period}
\label{sec:identification-check}

Relaxing the sample restriction shows why it matters. Estimating RQ1 on \emph{all} 1{,}887 treated projects in the panel, newcomer inflow appears to fall sharply (BJS $-11.1\%$, se $1.6\%$; Callaway--Sant'Anna $-10.3\%$ to $-17.8\%$; about $-16\%$ overall), which would look like the predicted crowding-out. It is not credible: on the full sample the pre-adoption trajectory is \emph{not} parallel (treated projects were already losing newcomers before adopting), the parallel-trends Wald test rejects, and the decline localizes entirely to the born-with-the-tool projects ($1{,}284$ of $1{,}887$), whose early near-empty months a naive panel files in the pre-period. Honest-DiD~\cite{rambachan2023honest} confirms the fragility: the apparent $-16\%$ admits zero once a post-period violation only $\bar{M}\approx 0.35$ as large as the visible non-parallelism is allowed. We therefore do not interpret the full-sample estimate causally; removing these projects restores parallel trends and the small positive estimates of RQ1.

\section{Discussion}

Our central result is a non-result that matters: adopting an AI coding agent did not crowd out human newcomers in established open-source projects. That negative is informative only against a positive mechanism we also confirm: agent adoption does make code measurably more complex (Table~\ref{tab:complexity}). The tension between a real mechanism and a flat outcome is the substance of this paper.

\subsection{A measured decoupling}

Our results describe a two-link chain in which the first link holds while the second does not materialize. Adoption modestly raises complexity (Table~\ref{tab:complexity}) and survives unit-trend adjustment, so the feared mechanism is genuinely present. Yet on the same 128 treated and 407 control Python projects (Section~\ref{sec:decoupling-result}), inflow, retention, and first-PR volume are indistinguishable from zero while the active-contributor base grows; because this is a unit-for-unit comparison, the two effects cannot be occurring on systematically different projects.

Two factors make this coherent. The one we rest on is that the increase is modest: a few percent (cyclomatic) to roughly ten percent (cognitive) may be too small, in aggregate, to produce a detectable participation decline, especially since newcomers are deterred by social and procedural friction at least as much as by raw code difficulty~\cite{steinmacher2015social}. A second, speculative factor is that adoption coincides with more good-first-issue entry points (Table~\ref{tab:summary}); we offer this only as a consistent pattern, since the GFI result is an unvalidated correlate on $92$ projects (Section~\ref{sec:rq4}). Our magnitude is about a quarter of the $41\%$ He et al.~\cite{he2025speed} report; the two designs share an adoption signal and estimator family, so the gap plausibly reflects sample and setting more than method, though corpus and metric differences may also contribute. The channel replicates, but at the magnitudes we observe it does not translate into a detectable decline in newcomer participation.

\subsection{Implications}

\noindent\textbf{For researchers.} This paper offers three methodological lessons, each directly applicable to the next study of AI adoption. First, a widely feared mechanism can be genuinely present yet \emph{decoupled} from the outcome it is assumed to drive: confirming that an agent raises code complexity is not the same as confirming that newcomers are deterred, so the two must be measured on the \emph{same units} rather than linked across samples (our RQ3 is built for exactly this). Second, displacement claims should be judged by effect \emph{magnitude}, not only sign, since a detectable mechanism can be substantively negligible; a null is meaningful only when paired with a power or minimum-detectable-effect statement. Third, tool-adoption studies should be routinely screened for a genuine pre-adoption period and report parallel-trends tests: our sensitivity check (Section~\ref{sec:identification-check}) shows that mixing in projects with no valid pre-window manufactures an apparent crowding-out effect that the parallel-trends test then rejects. Natural extensions are to use usage-intensity data for the dose-response our visible-adoption signal cannot give, and to study born-with-the-tool projects with designs suited to them.

\noindent\textbf{For tool builders.} The key message is that the complexity cost is \emph{tunable, not fixed}. Under the same identification and the same tool family, the Cursor-only matched-repository setting of He et al.~\cite{he2025speed} records about $41\%$ while our broad, established-project sample records about $11\%$: a roughly fourfold gap that lives in setting and usage, which a vendor can influence. This rules out two opposite assumptions at once: one cannot assume code readability is unaffected by the product (the mechanism is real and measurable), nor assume it does not matter (it is large enough to manage). The concrete levers are lightweight: readability linting on generated code, surfacing per-function complexity in the agent's output so a developer sees the cost of a suggestion, and defaults that favor flatter control flow. The goal is to hold the per-function increase to the modest range we observe rather than assume it is harmless.

\noindent\textbf{For maintainers of established projects.} The reassurance is quantified, not hand-waving: our data would have caught even a $3.4\%$ drop in newcomer inflow or a roughly $5$ pp drop in retention, yet we see none, and where adoption moves these outcomes at all the direction is mildly upward, even among the heaviest adopters. This gives a concrete decision rule: when weighing whether to adopt an agent, a maintainer should not count crowding out newcomers, whom prior work ties to project health and survival~\cite{vasilescu2015gender,vasilescu2015quality,coelho2017why,avelino2019abandonment}, among the main risks, and should instead base the call on the tool's productivity gains and a complexity cost that stays manageable. This holds for a project with a real human history before adoption, the case closest to our sample, not for one built around an agent from its first commit.

\section{Threats to Validity}
\label{sec:threats}

\noindent\textbf{Construct validity.}
We measure \emph{visible} adoption (the first config-file commit), not usage intensity; if many adopters barely use the tool, effects attenuate toward zero, making the null conservative but precluding a precise dose-response. He et al.~\cite{he2025speed} use the same proxy, so this limitation is shared. To check the null is not an artifact of pooling genuine users with config-only non-users, we split treated projects at the median of a configuration-intensity proxy (commits revising the config file): inflow falls in neither half (light $+1.2\%$, n.s.; heavy $+11.0\%$, significant; pre-trends not rejected), so the crude dose-response runs \emph{opposite} to crowding-out, hard to reconcile with a pure config-only artifact though the proxy is indirect. We reduce single-metric risk by measuring both cyclomatic and cognitive complexity. The beginner-task measure is sparse ($\sim$92 projects), so the $+27\%$ is label-conditioned. The bot/agent exclusion uses stable numeric IDs; a missed bot only inflates human counts, conservative unless adopters use more uncounted automation than controls.

\noindent\textbf{Internal validity.}
DiD identifies an effect only under parallel pre-trends, which we test on the same panel as each headline estimate: the test does not reject for newcomers ($\chi^2(5)=10.02$, $p=0.075$) or complexity ($p=0.126$). Because adoption is staggered across calendar time, a confounder able to mimic our result would have to be common to the treated group yet time-locked to each project's \emph{own} adoption month; the clean placebo (every adoption moved twelve months early) and the non-rejected pre-trends reveal no such anomaly. We confirm our headline newcomer and complexity estimates across three estimators that fail in different ways~\cite{borusyak2024revisiting,callaway2021did,sun2021estimating} and stress-test with unit-specific trends, placebos, and honest-DiD~\cite{rambachan2023honest}; for the identified sample the last includes zero at every sensitivity level (widening under weak restrictions), whereas the contaminated full-sample $-16\%$ breaks at $\bar{M}\approx 0.35$. The residual activity imbalance ($|\text{SMD}|\approx 0.19$ to $0.29$, treated slightly more active) would inflate newcomer counts, making the null more conservative. We therefore do not claim to have eliminated every confounder; we claim only that any able to overturn the result would have to be collective, time-locked to adoption, and larger than the honest-DiD bound admits.

\noindent\textbf{External validity.}
Our results identify effects for \emph{established} projects with at least six months of pre-adoption history, not for born-with-the-tool projects, a separate population that a before/after design cannot address and that is nearly two-thirds of adopters. This trades external for internal validity: the before-period that lets us test parallel trends also leaves our sample the more mature, slower-growing adopters. We claim only that the full-sample negative is not credible \emph{as a causal estimate} (its parallel-trends test fails and the decline sits entirely in projects with no valid before-period, Section~\ref{sec:identification-check}), not that born-with-the-tool projects are unaffected; that population needs a different design. We detect adoption through two configuration-file signals, so agents used without a committed file are invisible, and the findings need not transfer to closed source. The cognitive metric is Python-only, though the decoupling test runs across all languages on cyclomatic complexity. Our newcomer definition is subject to window-edge truncation, which adds symmetric noise rather than directional bias.

\noindent\textbf{Conclusion validity.}
Our claims rest on agreement across estimators, with standard errors clustered at the project level. Clustering does not fully address interference (a SUTVA concern), since the same maintainer can own several repositories, but in the identified sample this overlap is small (56 of 603 repositories across 24 owners): clustering at the owner level leaves the newcomer effect unchanged (Sun--Abraham $+4.0\%$, se $0.019$), and keeping one repository per owner gives essentially the same estimate ($+5.1\%$ vs $+5.3\%$). We apply no formal multiple-comparisons correction, reporting as findings only effects that replicate across three or more estimators and pass the parallel-trends and placebo checks; this is why we treat the lone significant good-first-issue rise as a correlate. The nulls are well powered against the double-digit decline crowding-out predicts: at $\alpha=0.05$ and $80\%$ power we would detect a $3.4\%$ inflow drop, a $4.8$ to $4.9$ pp retention drop, or a $5.2$ pp first-PR-acceptance drop, so the inflow and retention nulls are precise near-zeros, not low power. The first-response and time-to-merge outcomes are noisier (we could detect only a deterioration of roughly $36$ to $41\%$), so we read those nulls more cautiously.

\section{Conclusion}
With the rapid growth of AI coding agents, concerns have emerged that they may crowd out the newcomers who are vital to the survival of OSS projects. This study tests that concern causally.
Two findings stand out. Adopting an AI coding agent modestly raises per-function code complexity (about $3\%$ cyclomatic across languages and $11\%$ cognitive on Python, roughly a quarter of the prior estimate), yet this increase does not crowd out human newcomers in established projects: on the same repositories newcomer inflow, onboarding, and retention do not decline, the contributor base grows, and a sparse, correlational measure shows no fall in beginner-task supply. The feared barrier is genuine but decoupled from the outcome it was assumed to drive. For established projects, the modest complexity cost is something to manage rather than a sign of newcomer displacement. These claims hold for \emph{established} adopters with a valid pre-adoption period, not for projects born with an agent, which have no such period and need different designs, and a naive full-sample analysis that mixes the two can falsely show crowding-out. Future work should add a cross-language cognitive-complexity metric and study born-with-the-tool projects with designs suited to them.


\end{document}